\documentclass[12pt,dvips,a4paper]{article}
\usepackage{epsfig}
\include{commands}
\usepackage{amssymb}
\usepackage{amsmath}
\usepackage{setspace}
\usepackage[numbers]{natbib}
\usepackage{lineno}

\renewcommand{\title}{Analysis of the microbond test using nonlinear fracture mechanics}

\begin{document}

\begin{center} \begin{LARGE} \textbf{\title} \end{LARGE} \end{center}

\begin{center} 
J. Nov\'{a}k, C. J. Pearce and P. Grassl$^*$\\
Department of Civil Engineering, University of Glasgow, Glasgow, UK

L. Yang and J. Thomason\\
Department of Mechanical Engineering, University of Strathclyde, Glasgow, UK

$^*$ Corresponding author: grassl@civil.gla.ac.uk

Submitted to 17th international conference on composite materials (ICCM-17)

Last modified: 28th of May 2009

\end{center}

\begin{center}\textbf{SUMMARY}\end{center}
Microbond tests composed of single fibre and matrix droplet are often used to determine the properties of fibre reinforced composites.
Interfacial shear strength is quantified by the maximum pull-out force assuming a uniform stress distribution along the fibre.
Here, nonlinear finite element analyses are performed to investigate the validity of this assumption.

\textit{Keywords: Microbond test, Fibre reinforced composite, Nonlinear finite element analysis}

\begin{center}\textbf{INTRODUCTION}\end{center}
When determining the interfacial shear strength by the microbond test, it is commonly assumed that the relationship between pullout force and embeddment length is linear. 
However, many experimental results are in disagreement with this assumption \cite{PisDutLau98,LiuNair99}, which might be explained by the influence of boundary conditions, by the presence of thermal stresses and by the fracture propagation along the fibre.
In this work, these possible effects are investigated for microbond systems consisting of a Polypropylene (PP) matrix and a glass fibre.

The behaviour of fibre reinforced composites is strongly influenced by the interface properties between fibre and matrix.
One of the commonly used methods to determine these properties is the microbond test.
In this test, a fibre embedded in a droplet of matrix material is pulled through a gap formed by two knife blades, so that the droplet is sheared off the fibre \cite{MilMurReb87,GauMil89}. 
From this test, the average shear strength (ITTZ) is determined as the maximum force that is carried by the fibre divided by the area of embeddment of the fibre.  This average shear strength is then used for the modelling of the response of fibre reinforced composites consisting of many fibres.

Although the concept of the microbond test is simple, its loading setup results in a complex nonuniform shear stress distribution along the fibre, which changes during the pull-out process. 
Phenomena influencing this shear stress distribution are the presence of radial stresses and the fracture process during pullout. 
Radial stresses are caused by the knife blades pressing against the droplet and the nonuniform geometry of the droplet itself. In addition, radial stresses are also generated by the difference in thermal contraction of fibre and matrix during the manufacturing process of the test setup, which involves cooling down of the fibre and the melted matrix.

In this study, the microbond test is analysed by a three-dimensional finite element model.
The fracture process is described by nonlinear fracture mechanics (NLFM). 
NLFM describes the nonlinear response within the fracture process zone by means of softening stress-strain relationships.
For local constitutive models, the stress at a point depends on the history of this point only.  
Fracture is represented by localised strain profiles within finite elements, whereby the integral of the energy dissipation density over the size of the finite element results in the dissipated energy. 
Thus, for local softening constitutive models for fracture, the dissipated energy depends on the size of the finite element.
However, the energy dissipated during the fracture process is a material property and should be independent of the discretisation applied.
This limitation of local constitutive models can be overcome by adjustment of the softening modulus of the softening branch of the stress-strain curve with respect to the finite element size \cite{Pietruszczak81,BazOh83,WilBicStu86}.
With this approach, the dissipated energy is modelled mesh independently, as long as the inelastic strains localise in a zone of assumed size.
This approach, which was used earlier for the analysis of delamination in sandwich structures \cite{BazGra07} and cracking in cohesive materials \cite{JirGra08}, is used in the present study.

The present NLFM approach differs strongly from linear elastic fracture mechanics (LEFM) approaches, which assume a large stress-free crack with all the nonlinearities of the fracture process concentrated in an infinitesimally small zone in front of the crack tip \cite{Gri21}.
If the size of the fracture process zone is large compared to the length of the stress free crack and the size of the structure, LEFM results in a poor approximation of the fracture process. It is believed  that this is the case for the polyprolyene matrix system. Therefore, nonlinear fracture mechanics (NLFM) is chosen, which provides a more accurate description of the fracture process in these situations.

The aim of this work is to investigate the influence of radial stresses and the fracture process on the average shear stress obtained from the microbond test.
A parameter study is carried out to quantify, which of the influences is dominant for the case of the microbond test.
To the authors' knowledge, nonlinear fracture mechanics has not been used before to analyse the microbond test.
It is hoped that with this study, further insight into the microbond test is gained. 

\begin{center}\textbf{FINITE ELEMENT MODEL}\end{center}

The present section deals with the description of the geometry, boundary conditions and material laws used in the finite element modelling of the microbond test. 

\begin{center}\textbf{Droplet geometry and boundary conditions}\end{center}

\begin{figure}
\begin{center}
\begin{tabular}{cc}
\epsfig{file=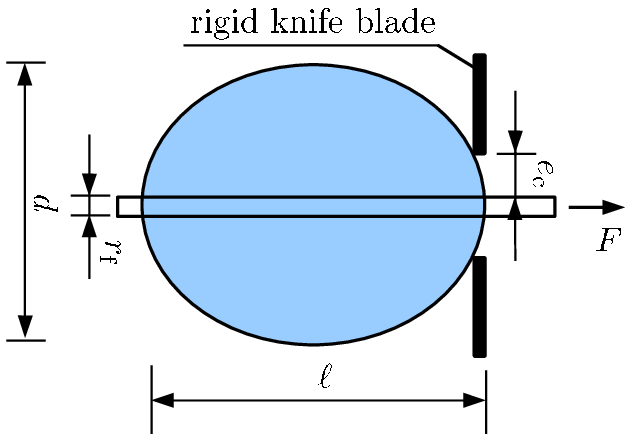, width = 6.8cm} &
\epsfig{file=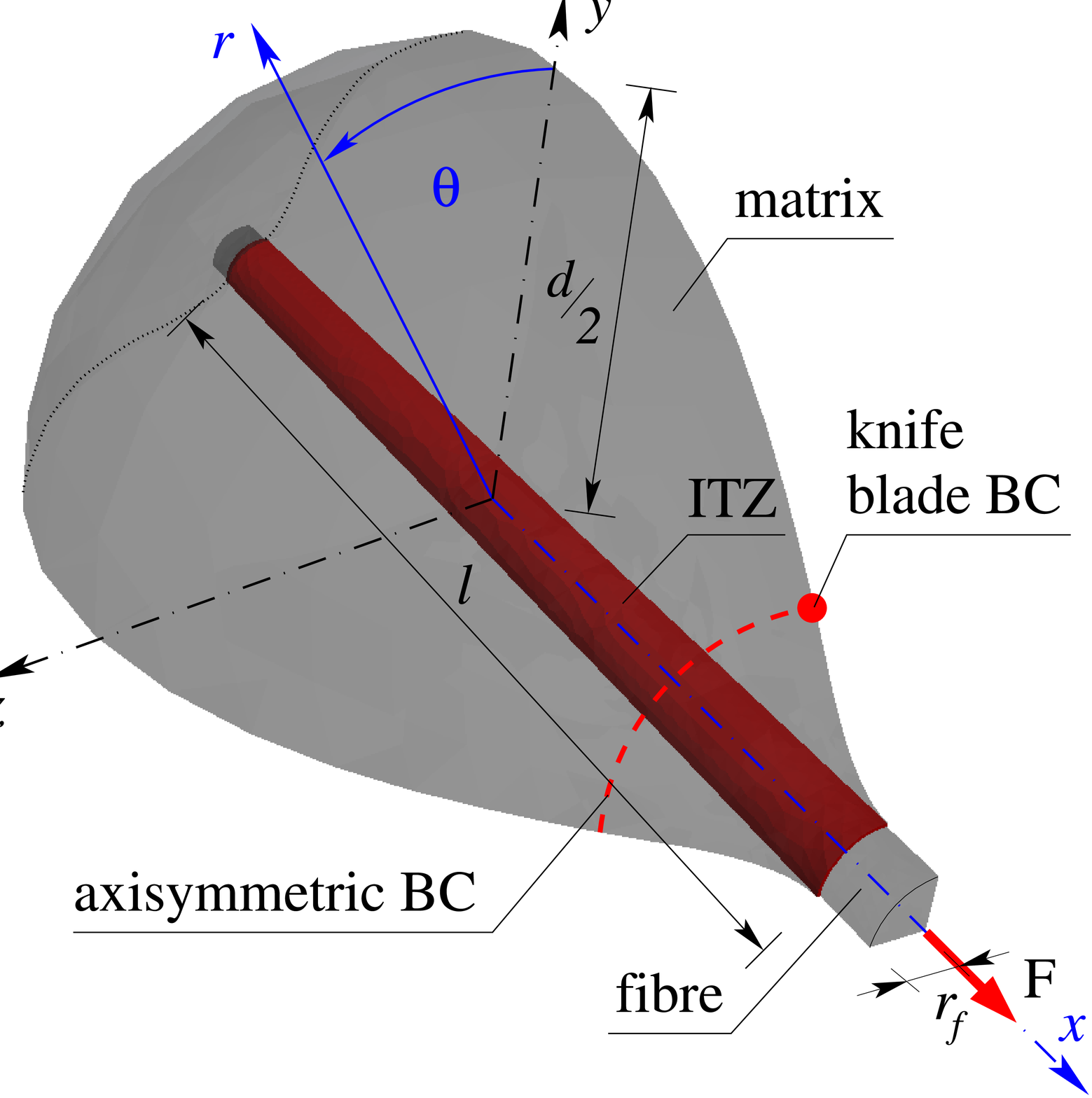, width = 6.5cm} \\
(a) & (b)
\end{tabular}
\caption{a)~Test setup of the microbond test. b) Geometry of the finite element model.} 
\label{fig:geom}
\end{center}
\end{figure}

The shape of the droplet is governed by the way it is manufactured. This involves the heating of a small amount of matrix material, attached to the fibre, until a droplet of the melted matrix material forms around the fibre.
The shape of the droplet is mainly determined by attractive forces between matrix and fibre, since the volume of the droplet is so small that gravitational effects are negligible.
The droplet geometry is derived by the Laplace excess pressure across the droplet surface as discussed in \cite{Car76, MenLetMan02}.
The geometry of the fibre droplet composite (Fig.~\ref{fig:geom}a~and~b) is determined by the fibre radius $r_{\rm f}$, the droplet diameter $d$ and the embedded length $\ell$.
In the finite element model shown in Fig.~\ref{fig:geom}b, only a quarter of the droplet is modeled. 
Furthermore, an interfacial transition zone (ITZ) of thickness $h_{\rm e}$ is introduced. 
This three-dimensional representation allows one to consider the influence of the knife blade.
This representation is advantageous over axisymmetric finite element models, which automatically assume axisymmetric boundary conditions.
In the literature, results of modified microbond tests have been reported, where the fibre is pulled through a circular hole in a plate \cite{PisDutLau98}.
This modified setup results in boundary conditions which are closer to axisymmetric.

The fibre end is subjected to a force $F$ in $x$-direction. 
The constraint of the droplet in the $x$-direction due to the knife blade is modeled by a kinematic boundary condition in one node (Fig.~\ref{fig:geom}b). The gap between the knife blade and the fibre is denoted as $e_{\rm c}$.
In the nonlinear finite element analysis, the pull-out force was controlled by an indirect displacement control using the relative displacement between the droplet and the fibre at the boundaries of the interfacial transition zone. 

One of the key points of this study is the influence of the embedded length on the interface strength and how this parameter is affected by radial stresses and the fracture process zone.
To be able to investigate this effect, three different droplet sizes were analysed with their dimensions presented in Tab.~\ref{tab:geometry}.
\begin{table}
\begin{center}
\caption{Model parameters of the structural analyses. All the values are given in $\mu$m.}
\label{tab:geometry}
\begin{tabular}{cccccc}
\hline
Droplet size & $\ell$ & $d$  & $r$  & $h_{\rm e}$  & $e_{\rm c}$ \\
Small & 100 & 27 & 8 & 0.5 & 10 \\
Medium & 215 & 77 & 8 & 0.5 & 10 \\
Large & 400 & 164 & 8 & 0.5 & 10\\
\hline
\end{tabular}
\end{center}
\end{table}
In the analyses for the three specimen sizes, the fibre radius $r_{\rm f}$ and the knife gap $e_{\rm c}$ were kept constant for all three specimen sizes (Tab.~\ref{tab:geometry}).

\begin{center}\textbf{Constitutive model}\end{center}

The material response of the fibre and the droplet is described by small strain isotropic linear elasticity with the stress-strain relationship
\begin{equation}
\boldsymbol{\sigma} = \mathbf{D}:\boldsymbol{\varepsilon}
\end{equation}
where $\boldsymbol{\sigma}$ is the stress tensor, $\boldsymbol{\varepsilon}$ is the strain tensor and $\mathbf{D}$ is the isotropic linear elastic stiffness tensor based on the Young's modulus $E$ and Poisson's ratio $\nu$.

The nonlinear interaction between the droplet and the fibre in the ITZ is modelled by an isotropic elasto-damage model in a thin layer of continuum elements in the droplet adjacent to the fibre.
The isotropic continuum model has the stress-strain relationship
 \begin{equation}\label{eq:stressStrain}
  \boldsymbol{\sigma} = (1-\omega) \mathbf{D}:\boldsymbol{\varepsilon}
   = \left(1-\omega\right)\bar{\boldsymbol{\sigma}}
 \end{equation}
Here, $\omega$ is the damage variable and $\bar{\boldsymbol{\sigma}}$ is the effective stress tensor, i.e. the stress that is carried by the undamaged material.
The damage variable $\omega$ is a function of history variable $\kappa$ which is defined as the maximum equivalent strain $\tilde{\gamma}$ reached in the history of the material: 
$ \kappa(t) = \max \tilde{\gamma}(\tau) $
for $\tau \leq t$, where $t$ is the time. 

\begin{figure}
\begin{center}
\begin{tabular}{cc}
\epsfig{file=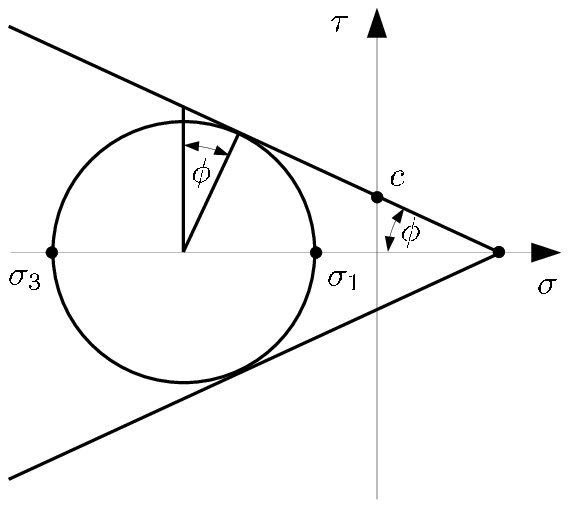, width = 6cm} &
\epsfig{file=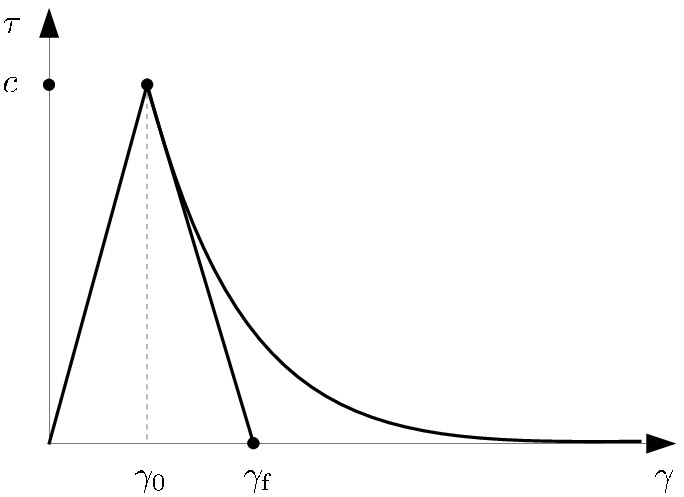, width = 7.cm} \\
(a) & (b)
\end{tabular}
\caption{a) The initial stress envelope represented in the principal stress
space obtained with the damage loading function. b) The softening stress-strain curve obtained with the damage model.}
\label{fig:constLaw}
\end{center}
\end{figure}

The equivalent strain is defined so that the Mohr-Coulomb strength envelope (Fig.~\ref{fig:constLaw}a) of the form
\begin{equation} \label{eq:Mohr}
|\tau| + \sigma \tan \phi - c = 0
\end{equation}
is obtained for $\tilde{\gamma} = \gamma_0 = (c \cos \phi)/G$, where $c$ is the cohesion, i.e the shear resistance at zero normal stress $\sigma$, $G$ is the shear modulus and $\phi$ is the angle of internal friction.

According to Fig.~\ref{fig:constLaw}a and \cite[chap. 15.1.5]{JirBaz01}, the strength envelope in Eq.~(\ref{eq:Mohr}) can be written as
\begin{equation}
\dfrac{\sigma_1 - \sigma_3}{2} = \left(c - \dfrac{\sigma_1+\sigma_3}{2}\tan \phi \right) \cos \phi
\end{equation}
This gives,
\begin{equation} \label{eq:MohrMod}
\left(1+\sin \phi\right)\sigma_1 - \left(1 -\sin \phi\right) \sigma_3 = 2 c \cos \phi
\end{equation}
For a pure shear state, $\tau_{\rm m} = \sigma_1 = -\sigma_3$ and Eq.~(\ref{eq:MohrMod}) reduces to
\begin{equation} \label{eq:MohrSpecial}
2 \tau_{\rm m} = 2 c \cos \phi
\end{equation}
The equivalent strain in this pure shear stress state is chosen to be $\tilde{\gamma} = \tau_{\rm m}/G$.
For general stress states, the equivalent strain definition is determined by setting Eq.~(\ref{eq:MohrSpecial}) and Eq.~(\ref{eq:MohrMod}) equal and solving for $\tilde{\gamma}$. This gives,
\begin{equation}
\tilde{\gamma} = \dfrac{1}{2G} \left( \left(1+\sin \phi\right)\sigma_1 - \left(1 -\sin \phi\right) \sigma_3\right)
\end{equation}

The strength envelope in Eq.~(\ref{eq:Mohr}) turns into the Tresca strength envelope for $\phi = 0^{\circ}$ and into the Rankine strength envelope for $\phi=90^{\circ}$. For the Tresca strength envelope, a pure shear stress state is limited by the cohesion $c$ independent of normal stresses. For the Rankine criterion, for which the shear strength is strongly dependent on the normal stress, the strength is determined by the maximum principal stress component. These two limiting cases will be used to investigate the influence of thermal stresses on the response of the microbond test. For $\phi=0^{\circ}$, the interfacial strength given by the microbond test is expected to be independent of the thermal stress. However, for $\phi=90^{\circ}$ a strong influence is expected.

The damage variable $\omega$ is related to the history variable $\kappa$ as
\begin{equation}\label{eq:softeningFEA}
  \omega = g\left(\kappa\right) = \left\{ \begin{array}{ll}
  0 & \mbox{if $\kappa \leq \gamma_0$} \\
  1 - \dfrac{\gamma_0}{\kappa}\exp\left( - \dfrac{\kappa -
  \gamma_0}{\gamma_{\rm f} - \gamma_{\rm 0}}\right) &
  \mbox{if $\kappa \geq \gamma_0$}
 \end{array} \right.
\end{equation}
The parameter $\gamma_{\rm f}$ is related to the fracture energy $G_{\rm F}$ in shear as
\begin{equation} \label{eq:Gf}
  \gamma_{\rm f} = \dfrac{G_{\rm F}}{ E \gamma_0
  h_{\rm e} } + \dfrac{1}{2}\gamma_0
 \end{equation} %
where $h_{\rm e}$ is the depth of the element row (softening band) next to the fibre, which represents the ITZ. 
This damage law results in a softening stress-strain curve in shear, as presented in Fig.~\ref{fig:constLaw}b.
The fracture energy, which is expected to influence strongly the interfacial shear strength, is one of the parameters, which will be varied in the present study.
The material parameters of the three phases were chosen according to Tab.~\ref{tab:mat} for a temperature of $T = 20$~$^{\circ}$C.
\begin{table}
\begin{center}
\caption{Material parameters for the three phases at a temperature of $T = 20$~$^\circ$C.}
\label{tab:mat}
\begin{tabular}{cccccc}
  \hline
  Phase & $E$ [GPa] & $\nu$ & $c$ [MPa] & $G_{\rm c}$ [J/m$^2$] & $\alpha_{\rm T}$~[1/K] \\
  Fibre & 7.6 & 0.22 & - & - & $1 \times 10^{-6}$\\
  Matrix & 2.5 & 0.2 & - & - & $111 \times 10^{-6}$\\
  ITZ & 2.5 & 0.2 & 5 & 3000 & $1\times 10^{-6}$\\
  \hline
\end{tabular}
\end{center}
\end{table}

\begin{figure}
\begin{center}
\begin{tabular}{cc}
\epsfig{file=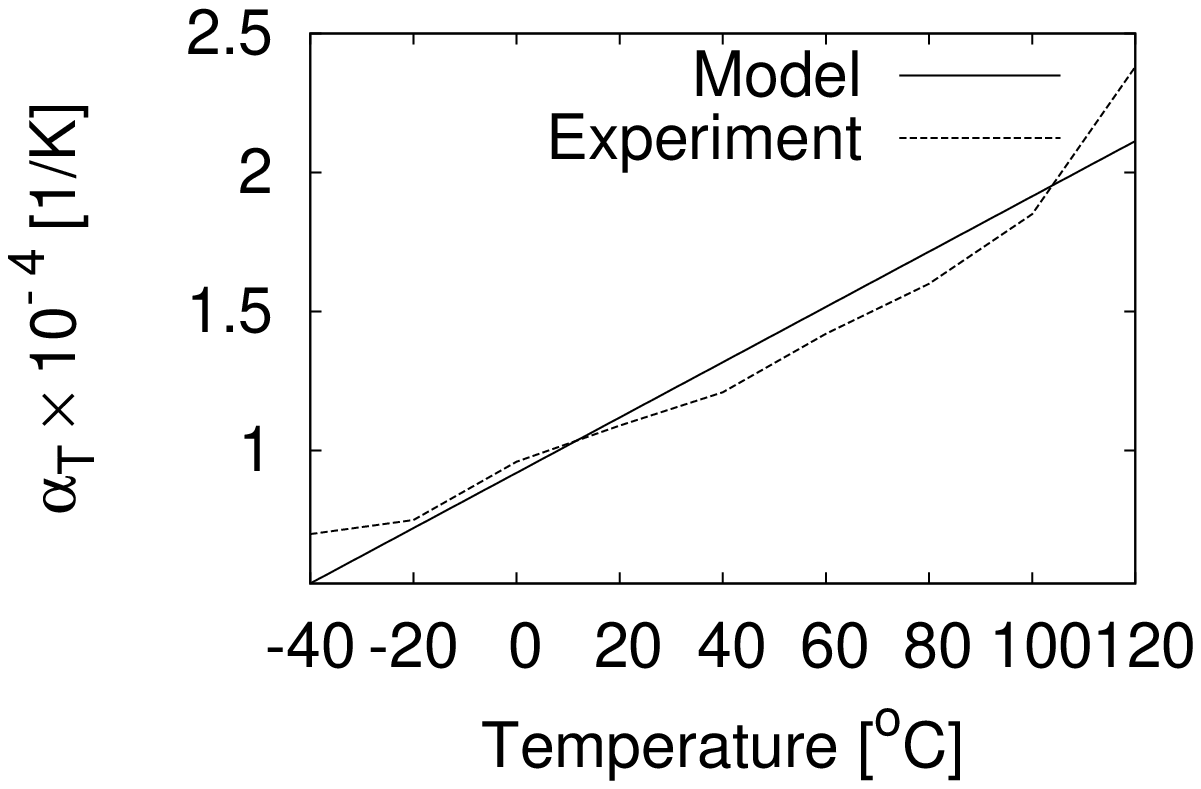, width=8cm} & \epsfig{file=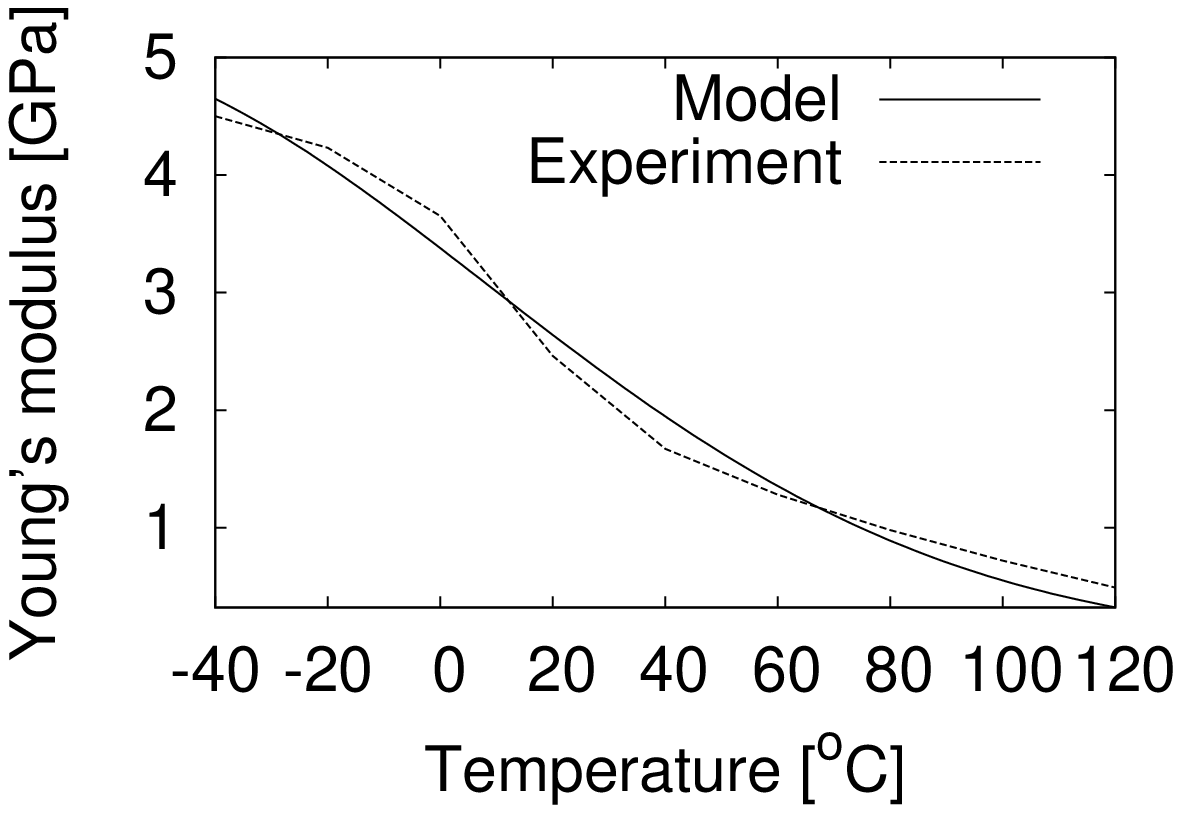, width=8cm}\\
(a) & (b)
\end{tabular}
\end{center}
\caption{Comparison of thermal properties of the matrix used in the constitutive model to experimental results reported in \cite{ThoGro96}. Influence of the temperature on (a) thermal expansion and  (b) Young's modulus.}
\label{fig:matrixResponse}
\end{figure}

The matrix material undergoes a phase change during the cooling down, which influences the mechanical and thermal properties. This is considered in the constitutive model by temperature dependent functions for the Young's modulus and the thermal expansion coefficient.
A nonlinear thermal expansion coefficient of the form
\begin{equation}\label{eq:thermalCoeff}
\alpha_{\rm T}\left(T\right) = a_{\rm t} T + b_{\rm t}
\end{equation}
was chosen.
Furthermore, the temperature dependent stiffness of the matrix was chosen as
\begin{equation}\label{eq:thermalStiff}
E \left(T\right) = E_{\rm 0} \exp(-((T+a_{\rm e})/b_{\rm e})^2)
\end{equation}
The parameters in Eqs.~(\ref{eq:thermalCoeff})~and~(\ref{eq:thermalStiff}) are determined by a best fit to experimental results reported in \cite{ThoGro96} as $a_{\rm t} = 9.9 \times 10^{-7}$, $b_{\rm t} = 9.2 \times 10^{-5}$, $E_0 = 5$~GPa, $a_{\rm e} = 75$~$^{\circ}$C and $b_{\rm e} = 75$~$^{\circ}$C (Fig.~\ref{fig:matrixResponse}).
For the fibre and ITZ, the thermal expansion coefficient and Young's modulus in Tab.~\ref{tab:mat} for $T=20$~$^{\circ}$C was assumed to be independent of the temperature.

\begin{center}\textbf{RESULTS}\end{center}
The results are divided into two parts. Firstly, the effect of radial stresses are investigated. 
Secondly, the influence of the fracture process zone is studied. 

\begin{figure}
  \begin{center}
    \epsfig{file = 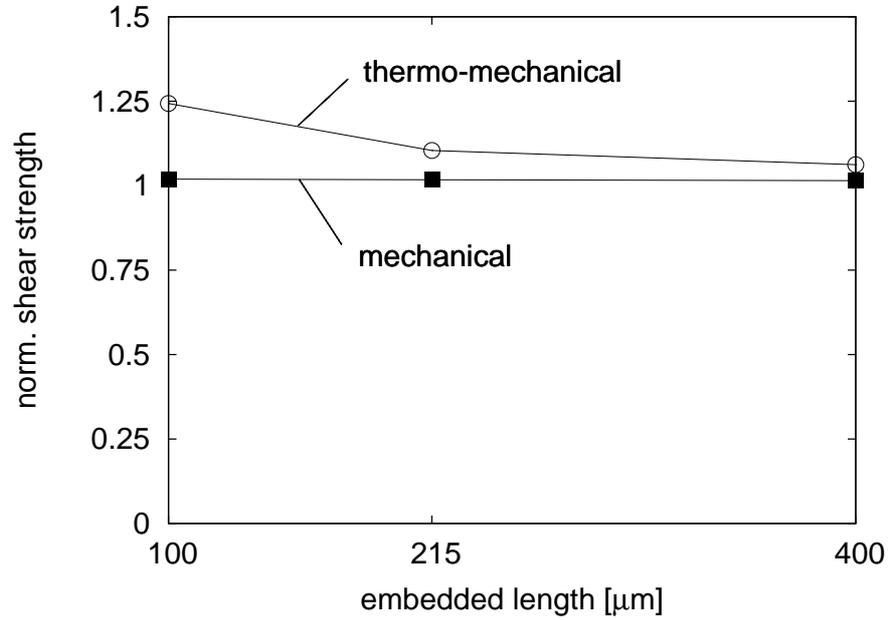,width=12cm}
    \caption{Normalised average shear stress versus embedded length for thermo-mechanical and mechanical analysis with the Rankine criterion ($\phi=90^{\circ}$).}
    \label{fig:resultsRadial}
  \end{center}
\end{figure}

Cooling down of the droplet is expected to generate large radial stresses. 
Furthermore, the pressure of the knife blades and the shape of the droplet will cause radial stresses as well. 
To be able to study the effect of thermal stresses, knife blades and geometry separately, analyses with cooling down and without cooling down are performed for the three droplet sizes presented in the previous section. 
For each of the six analyses, the average interfacial shear strength is determined by dividing the maximum pullout force by the embedded area of the fibre.
The Rankine criterion is used for the analyses, which results in the greatest increase of shear stresses due to radial compressive stresses.
The results of the analyses are presented in Fig.~\ref{fig:resultsRadial} in the form of the average shear stress normalised by the local shear strength $c$ (Tab.~\ref{tab:mat}).  

\begin{figure}
  \begin{center}
    \epsfig{file = 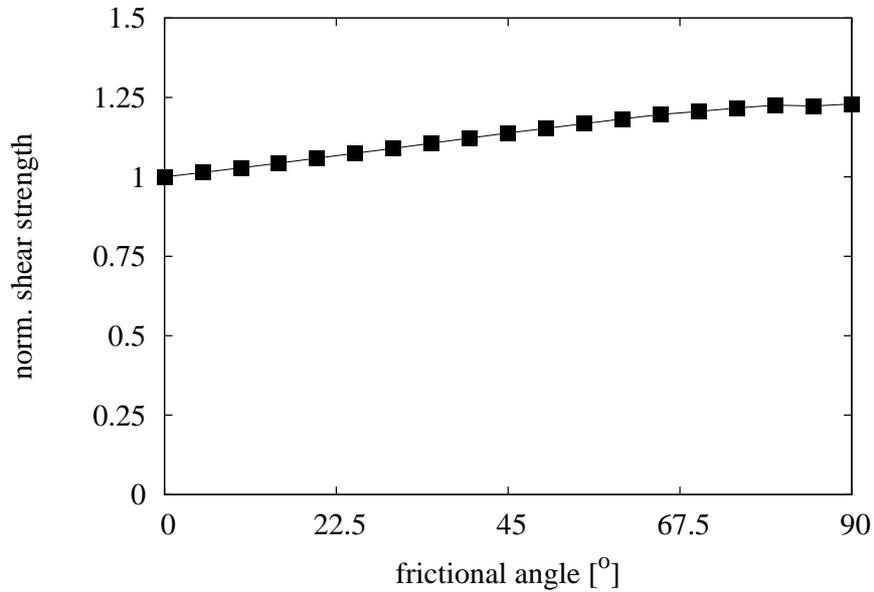,width=12cm}
    \caption{Normalised average shear strength versus frictional angle $\phi$ for thermo-mechanical analyses of the smallest specimen with an embedded length of $\ell = 100$~$\mu$m. The average shear strengths are normalised by a local shear strength of 5~MPa.}
    \label{fig:resultsAngle}
  \end{center}
\end{figure}

\begin{figure}
  \begin{center}
    \begin{tabular}{cc}
      \epsfig{file = 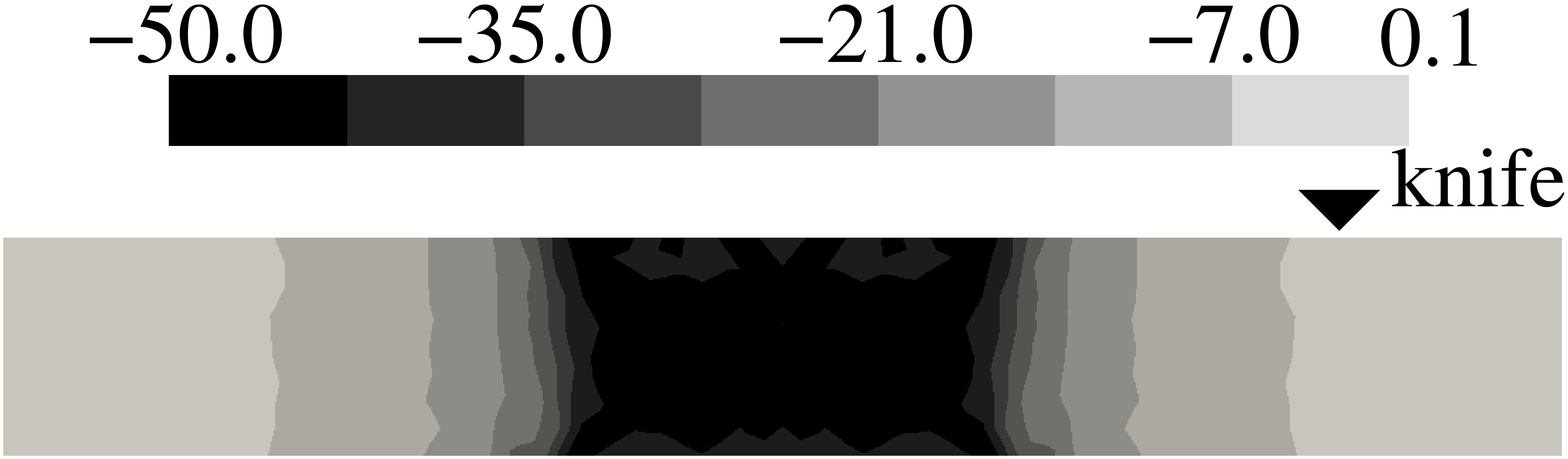,width=7cm} & \epsfig{file = 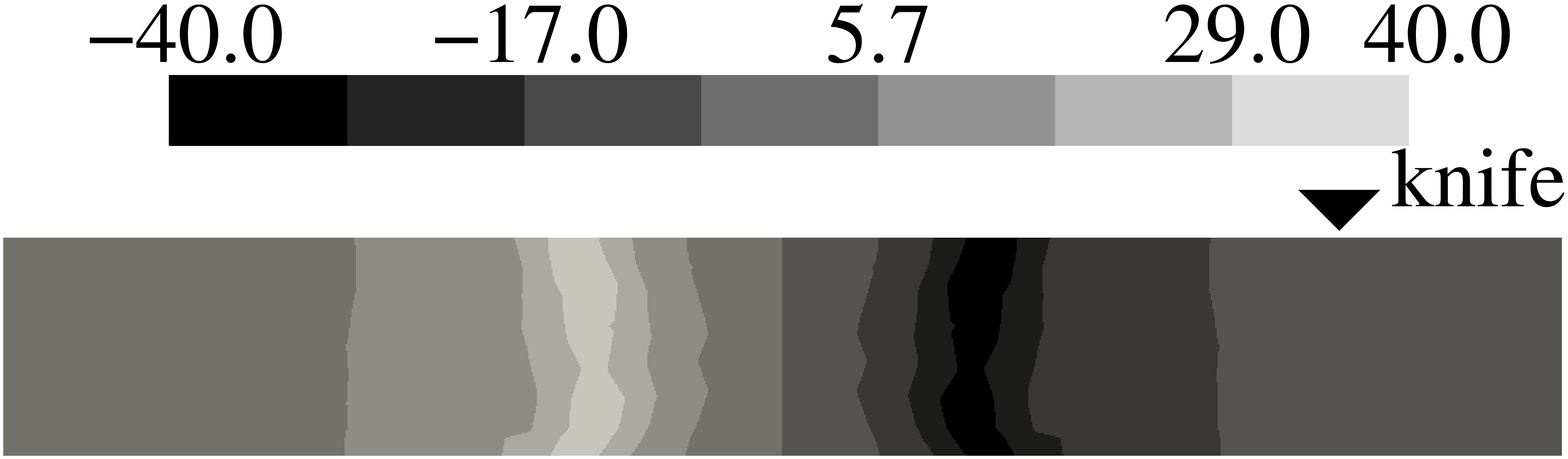,width=7cm}\\
      (a) & (b)
    \end{tabular}
    \caption{Contour plot of (a) radial stresses and (b) shear stresses in the ITZ after cooling down for a frictional angle of $\phi=90^{\circ}$ and a embedded length of $\ell = 100$~$\mu$m.}
    \label{fig:contourAngle1}
  \end{center}
\end{figure}

\begin{figure}
  \begin{center}
    \begin{tabular}{cc}
      \epsfig{file = 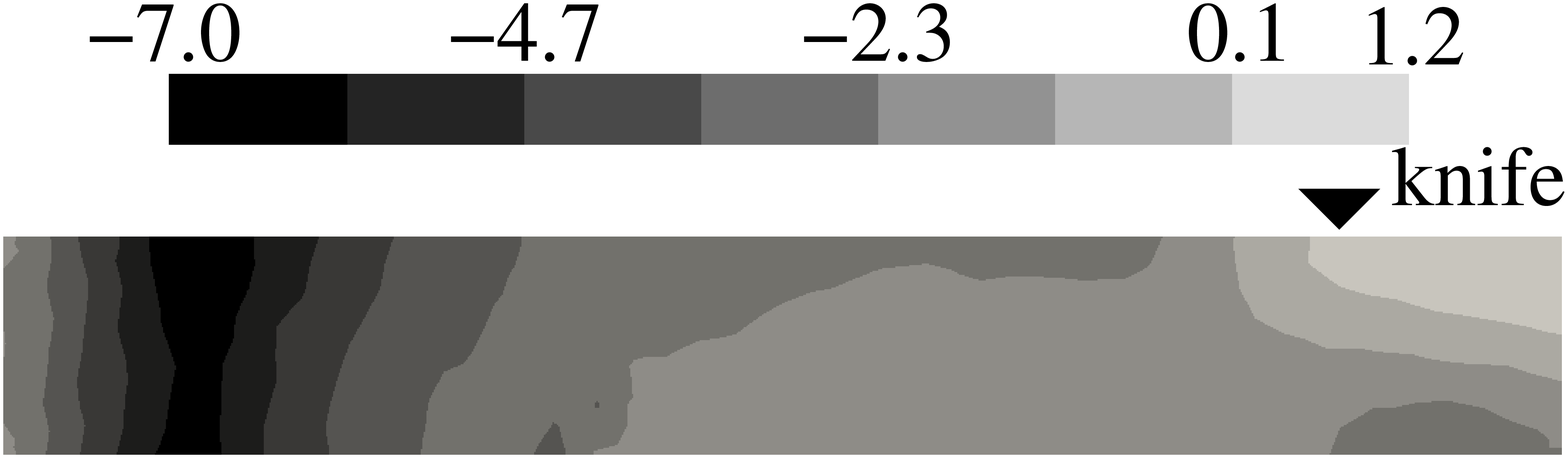,width=7cm} &  \epsfig{file = 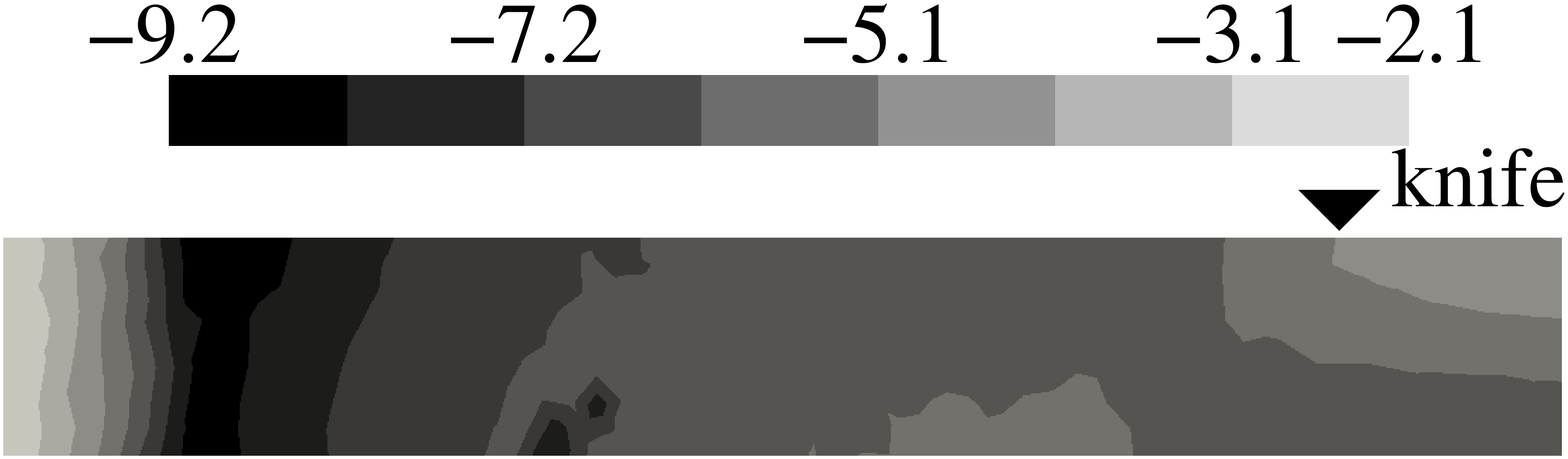,width=7cm}\\
      (a) & (b)
    \end{tabular}
    \caption{Contour plot of (a) radial stresses and (b) shear stresses in the ITZ at maximum pullout force for a frictional angle of $\phi=90^{\circ}$ and a embedded length of $\ell = 100$~$\mu$m.}
    \label{fig:contourAngle2}
  \end{center}
\end{figure}

For the thermo-mechanical analyses the average shear strengths are up to 25~\% greater than those obtained for the mechanical analyses alone. This indicates that the cooling down activates significant radial compressive stresses.
To investigate further the influence of thermally induced radial stresses and the role of the frictional angel $\phi$, a parameter study was performed, in which  thermo-mechanical analyses were carried out for the smallest specimen (100~$\mu$m) with the frictional angle varying from 0$^\circ$ to 90$^\circ$.
The results in the form of the normalised average shear strength versus the frictional angle $\phi$ is presented in Fig.~\ref{fig:resultsAngle}.
These results demonstrate that the frictional angle has a strong influence on the increase of shear stresses due to the radial compressive stresses. 
In experiments reported in \cite{YueQue92}, the frictional angle for polypropylene is reported to be in the range of 10$^\circ$ to 15$^\circ$. For an angle of friction of 15$^\circ$, the present analysis predicts the shear stress to be 6~\% greater than for $\phi=0^{\circ}$.
The influence of the cooling down is further investigated by studying the stress distribution in the ITZ along the fibre.
For a frictional angle of $\phi=90^{\circ}$ and an embedded length of $\ell=100$~$\mu$m, the radial and shear stresses (see Fig.~\ref{fig:geom}b for a definition of the cylindrical coordinate system) are presented in Fig.~\ref{fig:contourAngle1} after the cooling down and in Fig.~\ref{fig:contourAngle2} at the maximum pullout force.

After the cooling down (Fig.~\ref{fig:contourAngle1}), radial compressive stresses of up to 50~MPa are present. 
However, these stresses are reduced during the shearing off, so that the maximum radial compressive stress at maximum pullout force is only 7~MPa.

\begin{figure}
  \begin{center}
    \epsfig{file = 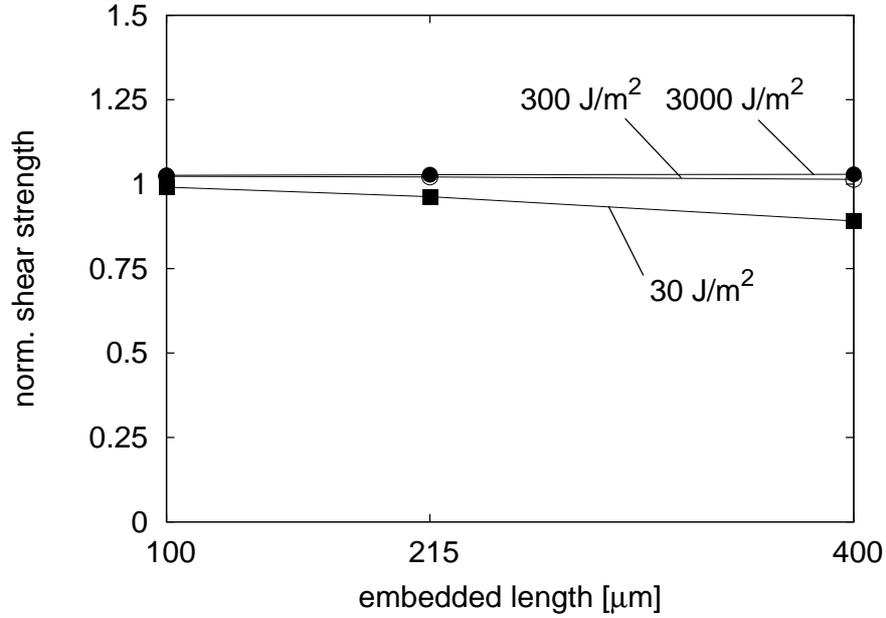,width=12cm}
    \caption{Normalised average shear stress versus embedded length for mechanical analysis with the Rankine criterion ($\phi=90^{\circ}$) for fracture energies $G_{\rm f} = 30,300$~and~$3000$~J/m$^2$.}
    \label{fig:resultsEnergy}
  \end{center}
\end{figure}

\begin{figure}
  \begin{center}
    \begin{tabular}{cc}
      \epsfig{file = 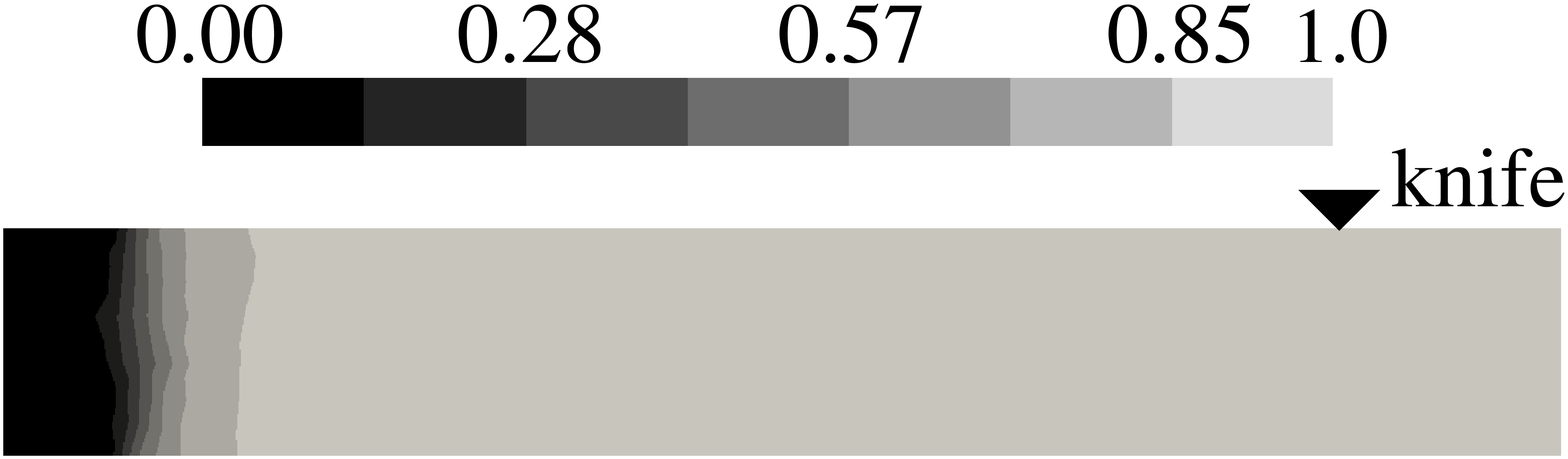,width=7cm} &  \epsfig{file = 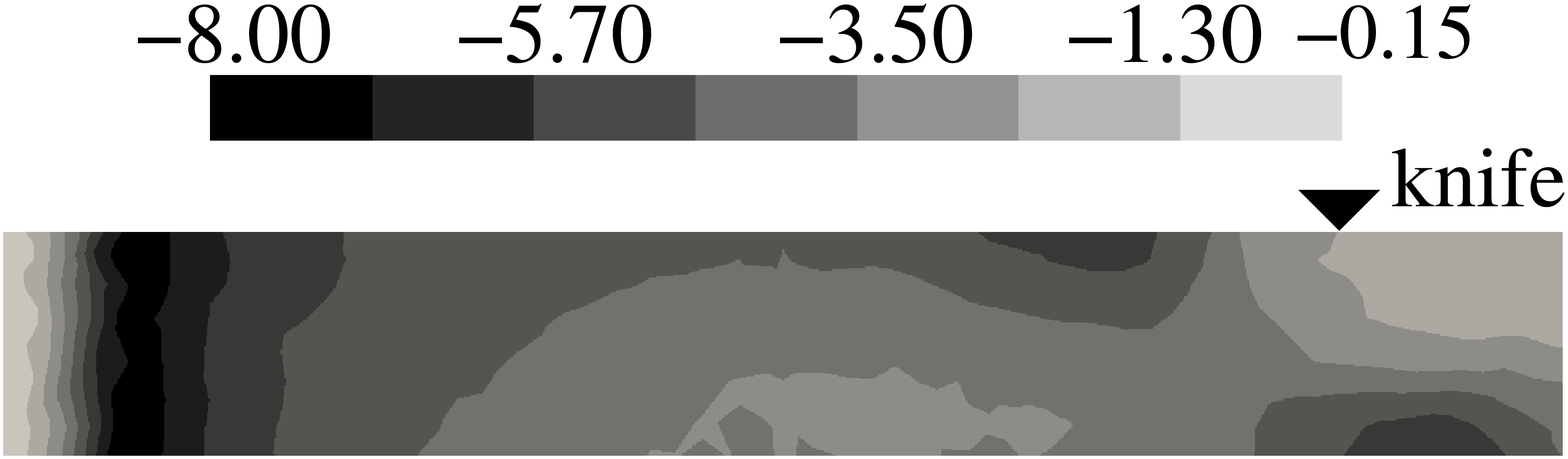,width=7cm}\\
      (a) & (b)
    \end{tabular}
    \caption{Contour plot of (a) damage parameter $\omega$ and (b) shear stresses in the ITZ at maximum pullout force for a frictional angle of $\phi=90^{\circ}$ and an embedded length of $\ell = 100$~$\mu$m for a mechanical analysis without the effect of cooling down.}
    \label{fig:contourEnergy}
  \end{center}
\end{figure}

In the second part of the analyses, the influence of the fracture energy is investigated. 
The responses of the three droplet sizes with $\ell=100$, $215$~and~$400$~$\mu$m are investigated for fracture energies $G_{\rm f} = 30$, $300$~and~$3000$~J/m$^2$. 
The pullout process is analysed for a constant temperature of $T = 20$~$^\circ$C without taking into account the cooling down from $T=120$~to~$20$~$^\circ$C. In this way, only the mechanical response is considered, so that the influence of the fracture energy can be separated from the effect of thermal stresses.
The results are presented in the form of the normalised average shear stress versus the embedded length in Fig.~\ref{fig:resultsEnergy}.
With decreasing fracture energy, the averaged shear strength decreases. Furthermore, the average shear strength decreases with increasing embedded length. 
The influence of the embedded length is most pronounced for a fracture energy of $30$~J/m$^2$.
However, this fracture energy is much less than fracture energies of PP reported in the literature \cite{Sim75, ReyCan00, WagNaiDet95}, which range between 300 and 3000~J/m$^2$.
The decrease of the average shear strength with decreasing fracture energy is explained by the damage distribution in the ITZ along the fibre, which is presented together with the shear stress distribution in Fig.~\ref{fig:contourEnergy}.

\begin{center}\textbf{CONCLUSIONS}\end{center}
In the present work, the influence of radial stresses and the fracture process on the average shear strength determined by the microbond test were investigated by nonlinear finite element analysis.
The work resulted in the following conclusions:
\begin{itemize}
\item Radial stresses generated by the knife blades and the droplet geometry have a small influence on the average shear strength.
\item Radial stresses, due to the initial cooling down of the matrix material, increased the average shear stress by up to 25~\%.
\item Fracture energy influences the average shear stress only for values which are significantly smaller than measured in experiments.
\end{itemize}

The modelling approach used in this work is based on many simplifications.
Material nonlinearities are limited to the ITZ between fibre and matrix. All materials are described by linear elasticity assuming small strains.
Furthermore, the pressure due to the knife blade is described by a kinematic constraint at a single node.
Nevertheless, the present study might help to design future experimental programmes and could provide useful guidance for the development of modelling approaches for fibre reinforced composites.

\begin{center}\textbf{ACKNOWLEDGEMENTS}\end{center}
Funding by the Glasgow Research Partnership in Engineering (GRPE) under project ``Multi-scale modelling of fibre reinforced composites'' is gratefully acknowledged. The authors would also like to express their gratitude to Dr. Bo\v{r}ek Patz\'{a}k of the Czech Technical University for kind assistance with his finite element package OOFEM (www.oofem.org) \cite{Pat99,PatBit01}.

\bibliographystyle{plainnat}

\bibliography{general}

\end{document}